\documentclass[preprint,showkeys,amsmath,amssymb]{revtex4} 
\usepackage{graphicx,psfrag}
\renewcommand{\d}{{\rm d}}
\newcommand{\e}{{\rm e}}
\newcommand{\eref}[1]{(\ref{#1})}
\usepackage{graphicx}
\begin{document}

\title{A hybrid of the optimal velocity and the slow-to-start models
and its ultradiscretization}
\altaffiliation{Accepted for publication in JSIAM Letters.}

\author{Kazuhito Oguma}
  \altaffiliation[Also at ]{Gunma National College of Technology, 
  580 Toriba, Maebashi, Gunma 371--8530, Japan}
  \affiliation{Department of Mathematical Engineering and Information Physics,
  Faculty of Engineering, The University of Tokyo, 7--3--1 Hongo,
  Bunkyo-ku, Tokyo 113--8656, Japan}

\author{Hideaki Ujino}
  \email{ujino@nat.gunma-ct.ac.jp}
  \affiliation{Gunma National College of Technology, 
  580 Toriba, Maebashi, Gunma 371--8530, Japan}

\begin{abstract}
Through an extension of the ultradiscretization for the optimal
velocity (OV) model, we introduce an ultradiscretizable traffic flow model,
which is a hybrid of the OV and the slow-to-start (s2s) models. 
Its ultradiscrete limit gives a generalization of a special case
of the ultradiscrete OV (uOV) model recently proposed
by Takahashi and Matsukidaira.
A phase transition from free to
jam phases as well as the existence of multiple metastable states
are observed in numerically obtained fundamental diagrams 
for cellular automata (CA), 
which are special cases of the ultradiscrete limit of the hybrid model.
\end{abstract}


\renewcommand{\thefootnote}{\arabic{footnote}}

\maketitle

\section{Introduction}

Studies on microscopic models for vehicle traffic
provided a good point of view on the phase transition
from free to congested traffic flow. Related
self-driven many-particle systems have attracted
considerable interests not only from engineers but
also from physicists~\cite{Chowdhury2000,Helbing2001}.
Among such models, the optimal velocity model~\cite{Bando1995},
which successfully shows a formation of ``phantom traffic jams''
in the high-density regime,
is a car-following model describing an adaptation to 
the optimal velocity that depends on the distance
from the vehicle ahead.

Whereas the OV model consists of ordinary differential equations (ODE),
cellular automata (CA) such as the Nagel--Schreckenberg model~\cite{Nagel1992},
the elementary CA of Rule 184 (ECA184)~\cite{Wolfram1986}, the Fukui--Ishibashi
(FI) model~\cite{Fukui1996} and the slow-to-start (s2s)
model~\cite{Takayasu1993} are extensively used in analyses of traffic flow.
Recently, Takahashi and Matsukidaira proposed a discrete OV (dOV) model,
which enables an ultradiscretization of the OV model~\cite{Takahashi2009}.
The resultant ultradiscrete OV (uOV) model includes both the ECA184 
and the FI model
as its special cases. However, the s2s effect remains to be included in their
ultradiscretization. The aim of this letter is to present
an ultradiscretizable hybrid of the OV and the s2s models.

\section{The OV model and the s2s effect}
Imagine many cars 
running in one direction on a single-lane highway. Let $x_k(t)$ denote
the position of the $k$-th car at time $t$. No overtaking is assumed
so that $x_k(t)\leq x_{k+1}(t)$ holds for arbitrary time $t$.
The time-evolution of the OV model~\cite{Bando1995} is given by
\begin{equation}
  \dfrac{\d v_k(t)}{\d t}=\dfrac{1}{t_0}\Bigl(
  v_{\rm opt}\bigl(\Delta x_k(t)\bigr)-v_k(t)\Bigr),
  \label{def:aOV}
\end{equation}
where $v_{k}:=\frac{\d x_k}{\d t}$ and $\Delta x_k:=x_{k+1}-x_k$
are the velocity of the $k$-th car and the interval between the
cars $k$ and $k+1$, respectively. A function $v_{\rm opt}$ and 
a constant $t_0$ represent an optimal velocity and 
sensitivity of drivers, or the delay of drivers' response,
in other words.

Since the current velocity and the current interval
between the car ahead determine the acceleration through the
time-evolution and the optimal velocity, we classify
the OV model~\eref{def:aOV} as 
the acceleration-control type (aOV). On the other hand,
the OV model of the velocity-control type (vOV) was proposed
in earlier studies of the car-following models~\cite{Newell1961},
\begin{equation}
  v_k(t)=v_{\rm opt}\bigl(\Delta x_k(t-t_0)\bigr).
  \label{def:vOV}
\end{equation}
Replacement of $t$ in the above equation \eref{def:vOV} with $t+t_0$ and
the Taylor series of $v_k(t+t_0)$ yield
\begin{align*}
  v_{\rm opt}\bigl(\Delta x_k(t)\bigr) & = v_k(t+t_0)
  = v_k(t)+\dfrac{\d v_k(t)}{\d t}t_0
      +\dfrac{1}{2}\dfrac{\d^2 v_k(t)}{\d t^2}t_0^2+\cdots,
\end{align*}
which is rewritten as
\[
  \dfrac{\d v_k(t)}{\d t}+\dfrac{1}{2}\dfrac{\d^2 v_k(t)}{\d t^2}t_0+\cdots
  =\dfrac{1}{t_0}\Bigl(v_{\rm opt}\bigl(\Delta x_k(t)\bigr)-v_k(t)\Bigr).
\]
Thus we note that 
the aOV model \eref{def:aOV} is given by
neglection of the higher derivatives in the Taylor series of the
vOV model \eref{def:vOV}. 
Though the aOV model is more common in the studies on vehicle traffic,
we shall concentrate on an ultradiscretizable hybrid
of the vOV and the s2s models. Thus we call 
the vOV model \eref{def:vOV} simply as the OV model, hereafter.

Note that the input to 
the OV function~$v_{\rm opt}(x)$ in the OV model~\eref{def:vOV}
is the headway at a single point of time
$t-t_0$ that is prior to the present time $t$. Thus we may say
that the OV model describes, in a sense, ``reckless'' drivers
since the model pays no attention to the headway between
the time $t-t_0$ and the present time $t$. On the other hand,
``cautious'' drivers governed by the s2s 
model~\cite{Takayasu1993} keep watching and require enough length of
headway to go on for a certain period of time before they
restart their cars. The contrast between the two models suggests
the idea that the s2s effect and the OV model can be brought
together by appropriately choosing an effective distance
$\Delta_{\rm eff} x_k(t)$ containing information on the headway 
for a certain period of time going back from the present
as an input to the OV function~$v_{\rm opt}(x)$.
We shall see this idea works in what follows. 

What is crucial in the ultradiscretization of the aOV 
model~\cite{Takahashi2009} is the choice
of the OV function,
\begin{equation}
  v_{\rm opt}(x):=v_0\Bigl(
  \dfrac{1}{1+\e^{-(x-x_0)/\delta x}}
  -\dfrac{1}{1+\e^{x_0/\delta x}}\Bigr),
  \label{eq:Takahashi-Matsukidaira_OVF}
\end{equation}
where $v_0, x_0$ and $\delta x$ are positive constants.
In terms of the auxiliary functions,
\begin{align}
  & \widetilde{v}_{\rm opt}(x):=v_0\dfrac{\d\tilde{x}_{\rm opt}(x)}{\d x}
  \label{eq:aux1} \\
  & \widetilde{x}_{\rm opt}(x):=\delta x\log\bigl(1+\e^{(x-x_0)/\delta x}\bigr)
  \label{eq:aux2}
\end{align}
the OV function~\eref{eq:Takahashi-Matsukidaira_OVF} is expressed as
\begin{equation*}
  \label{eq:OVF}
  v_{\rm opt}(x)=\widetilde{v}_{\rm opt}(x)-\widetilde{v}_{\rm opt}(x=0).
\end{equation*}
A naive discretization of 
the auxiliary function~\eref{eq:aux1},
\begin{equation*}
  \label{eq:daux1}
  \widetilde{v}^{\rm d}_{\rm opt}(x):=
  \dfrac{\tilde{x}_{\rm opt}(x)-\tilde{x}_{\rm opt}(x-v_0\delta t)}{\delta t},
\end{equation*}
introduces 
the OV function for the discrete OV (dOV) model,
\begin{equation}
  \begin{split}
    & v^{\rm d}_{\rm opt}(x) = \widetilde{v}^{\rm d}_{\rm opt}(x)
    -\widetilde{v}^{\rm d}_{\rm opt}(x=0)
    = \dfrac{\delta x}{\delta t}\log
    \biggl[
      \dfrac{1+\e^{(x-x_0)/\delta x}}{1+\e^{-x_0/\delta x}}
      \bigg/\dfrac{1+\e^{(x-x_0-v_0\delta t)/\delta x}}
      {1+\e^{-(x_0+v_0\delta t)/\delta x}}
    \biggr],
  \end{split}
  \label{eq:dOVF}
\end{equation}
which is found to be ultradiscretizable.~\cite{Takahashi2009}

Let $x_k^n:=x_k(t=n\delta t)$ and $v_k^n:=(x_k^{n+1}-x_k^n)/\delta t$
where $n(=0,1,2\cdots)$ and $\delta t(>0)$ are the integral time
and the discrete time-step, respectively.
Employing the effective distance as
\begin{equation}
  \Delta_{\rm eff}^{\rm d}x_k^n:=\delta x\log\Bigl(\sum_{n^\prime=0}^{n_0}
  \dfrac{\e^{-\Delta x_k^{n-n^\prime}/\delta x}}{n_0+1}\Bigl)^{-1},
  \label{eq:discrete_effective_distance}
\end{equation}
where $n_0:=t_0/\delta t$,
we extend the OV model~\eref{def:vOV} in a time-discretized form as
\begin{equation}
  v_k^n=v_{\rm opt}^{\rm d}\bigl(\Delta_{\rm eff}^{\rm d} x_k^{n}\bigr),
  \label{eq:ds2s--OV}
\end{equation}
which is equivalent to
\begin{align*}
  x_k^{n+1}& =x_k^n+\delta x\Biggl\{
  \log\biggl[
      1+\Bigl(\sum_{n^\prime=0}^{n_0} 
      \dfrac{\e^{-(\Delta x_k^{n-n^\prime}-x_0)/\delta x}}{n_0+1}
      \Bigr)^{-1}\biggr]
  -\log\bigl(1+\e^{-x_0/\delta x}\bigr) \\
  & \quad -
  \log\biggl[
      1+\Bigl(\sum_{n^\prime=0}^{n_0} 
      \dfrac{\e^{-(\Delta x_k^{n-n^\prime}-x_0-v_0\delta t)/\delta x}}{n_0+1}
      \Bigr)^{-1}\biggr]
  +\log\bigl(1+\e^{-(x_0+v_0\delta t)/\delta x}\bigr)
  \biggr\}.
\end{align*}
It is straightforward to confirm
that the continuum limit $\delta t\rightarrow 0$ of the above
discrete s2s--OV (ds2s--OV) model~\eref{eq:ds2s--OV} reduces 
to the integral-differential equation
which we call the s2s--OV model,
\begin{equation}
  \dfrac{\d x_k(t)}{\d t} 
  = v_{\rm opt}\bigl(\Delta_{\rm eff} x_k(t)\bigr)
  = v_0\Bigl(1+\dfrac{1}{t_0}\int_0^{t_0}
  \e^{-(\Delta x_k(t-t^\prime)-x_0)/\delta x}\d t^\prime
  \Bigr)^{-1}
  -v_0\bigl(1+\e^{x_0/\delta x}\bigr)^{-1},
  \label{eq:s2s--OV}
\end{equation}
where the corresponding effective distance is given by
\begin{equation*}
  \Delta_{\rm eff} x_k:=\delta x\log\Bigl(\dfrac{1}{t_0}\int_0^{t_0}
  \e^{-\Delta x_k(t-t^\prime)/\delta x}\d t^\prime\Bigr)^{-1}.
  \label{def:effective_distance}
\end{equation*}
We shall see that the s2s effect is indeed built into the
OV model in the ultradiscrete limit of the ds2s--OV model. 

\section{Ultradiscretization}
Ultradiscretization~\cite{Tokihiro1996} is a scheme 
for getting a piecewise-linear equation from 
a difference equation via the limit formula
\[
  \lim_{\delta x\rightarrow +0}\delta x(\e^{A/\delta x}+\e^{B/\delta x}+\cdots)
  =\max(A,B,\cdots).
\]
In order to go forward to the ultradiscretization 
of the ds2s--OV model~\eref{eq:ds2s--OV},
it will be a good choice for us to begin with the ultradiscrete limit $\delta x\rightarrow +0$ of the auxiliary function~\eref{eq:aux2}:
\begin{equation}
  \widetilde{x}_{\rm opt}^{\rm u}(x)
  :=\lim_{\delta x\rightarrow +0}\widetilde{x}_{\rm opt}(x)=
  \max(0,x-x_0).
  \label{eq:uaux2}
\end{equation}
In the same way to make the OV function for the dOV model~\eref{eq:dOVF}
from the auxiliary function~\eref{eq:aux2},
we obtain the OV function for the uOV model~\cite{Takahashi2009} as
\begin{equation}
  v_{\rm opt}^{\rm u}(x) = \widetilde{v}_{\rm opt}^{\rm u}(x)
  -\widetilde{v}_{\rm opt}^{\rm u}(x=0)
  = \max\Bigl(0,\dfrac{x-x_0}{\delta t}\Bigr)
  -\max\Bigl(0,\dfrac{x-x_0}{\delta t} -v_0\Bigr),
  \label{eq:uOVF}
\end{equation}
where $\widetilde{v}_{\rm opt}^{\rm u}(x)
:=\bigl(\widetilde{x}_{\rm opt}^{\rm u}(x)
-\widetilde{x}_{\rm opt}^{\rm u}(x-v_0\delta t)\bigr)/\delta t$.
The effective distance~\eref{eq:discrete_effective_distance}, 
on the other hand, is ultradiscretized in the same manner:
\begin{equation}
  \Delta_{\rm eff}^{\rm u}x_k^n :=\lim_{\delta x\rightarrow +0}
  \Delta_{\rm eff}^{\rm d}x_k^n
  = -\max_{n^\prime=0}^{n_0}\bigl(-\Delta x_k^{n-n^\prime}\bigr)
  = \min_{n^\prime=0}^{n_0}\bigl(\Delta x_k^{n-n^\prime}\bigr).
  \label{eq:u_effective_distance}
\end{equation}
Thus we obtain an ultradiscrete equation
\begin{equation}
  v_k^n=v_{\rm opt}^{\rm u}\bigl(\Delta_{\rm eff}^{\rm u} x_k^{n}\bigr),
  \label{eq:us2s--OV}
\end{equation}
which is equivalent to
\begin{equation*}
  x_k^{n+1}= x_k^n+\max\Bigl(0,
  \min_{n^\prime=0}^{n_0}\bigl(\Delta x_k^{n-n^\prime}\bigr)-x_0\Bigr)
  -\max\Bigl(0,\min_{n^\prime=0}^{n_0}
  \bigl(\Delta x_k^{n-n^\prime}\bigr)-x_0-v_0\delta t\Bigr),
\end{equation*}
as the ultradiscrete limit of the ds2s--OV model~\eref{eq:ds2s--OV}.
We name it the ultradiscrete s2s--OV (us2s--OV) model.
When the monitoring period $n_0$ is fixed at zero, the us2s--OV model 
reduces to a special case of the uOV model~\cite{Takahashi2009}.
As we can see from eqs.~\eref{eq:uOVF}, \eref{eq:u_effective_distance} 
and \eref{eq:us2s--OV}, the velocity $v_{k}^n$ is determined
by the optimal velocity for the minimum headway in the period 
between $n-n_0$ and $n$. Thus cars will not restart nor
accelerate, unless enough clearance goes on for 
a certain period of time. On the other hand, cars immediately
stop or slow down when their headways become too small to keep
their velocities. The s2s effect and a ``cautious'' manner of driving
are built into the uOV model in this way.

Now let us see how a CA comes out from the us2s--OV model.
Let $x_0$ be the discretization step of the headway $\Delta x_k^n$,
or equivalently, the size of the unit cell of the CA. 
Then with no loss of generality, we may set $x_0=1$.
Assume that the number of vacant cells between 
the cars $k$ and $k+1$, $\widetilde{\Delta}x_k^n:=\Delta x_k^n -x_0$,
must be non-negative, $\widetilde{\Delta}x_k^n\geq 0$,
which prohibits car-crash.
Then the us2s--OV model~\eref{eq:us2s--OV} reduces to
\begin{equation}
  x_k^{n+1}=x_k^n+\min\Bigl(\min_{n^\prime=0}^{n_0}\bigl(\widetilde{\Delta}
  x_k^{n-n^\prime}\bigr),v_0\delta t\Bigr).
  \label{eq:s2s--OVCA}
\end{equation}
Fixing $v_0\delta t$ at an integer,
we call this model the s2s--OV cellular automaton (CA).
The s2s--OV CA reduces to the FI model~\cite{Fukui1996}
when $n_0=0$ and to the ECA184~\cite{Wolfram1986} 
when $n_0=0$ and $v_0\delta t=1(=x_0)$.
The s2s model~\cite{Takayasu1993} also comes out from the
s2s--OV CA by choosing $n_0=1$ and $v_0\delta t=1(=x_0)$.
Thus the s2s--OV CA is regarded as a hybrid of the FI model
and an extended s2s model.

\section{Numerical experiments}

We shall numerically investigate the s2s--OV CA~\eref{eq:s2s--OVCA}. 
Throughout this section, 
the length of the circuit $L$ is fixed at $L=100$ and 
the periodic boundary condition is assumed as well 
so that $x_k^n+L$ is identified with $x_k^n$.

\begin{figure}[h]
\psfrag{9}{\tiny 9}
\psfrag{8}{\tiny 8}
\psfrag{7}{\tiny 7}
\psfrag{6}{\tiny 6}
\psfrag{5}{\tiny 5}
\psfrag{4}{\tiny 4}
\psfrag{3}{\tiny 3}
\psfrag{2}{\tiny 2}
\psfrag{1}{\tiny 1}
\psfrag{0}{\tiny 0}
\psfrag{T}{\small Time}
\psfrag{i}{}
\psfrag{m}{}
\psfrag{e}{}
\psfrag{P}{\small\hspace*{-2em} Positions of Vehicles}
\psfrag{o}{}
\psfrag{s}{}
\psfrag{i}{}
\psfrag{t}{}
\psfrag{i}{}
\psfrag{o}{}
\psfrag{n}{}
\psfrag{s}{}
\psfrag{o}{}
\psfrag{f}{}
\psfrag{V}{}
\psfrag{e}{}
\psfrag{h}{}
\psfrag{i}{}
\psfrag{c}{}
\psfrag{l}{}
\psfrag{e}{}
\psfrag{s}{}
\includegraphics[width=60mm]{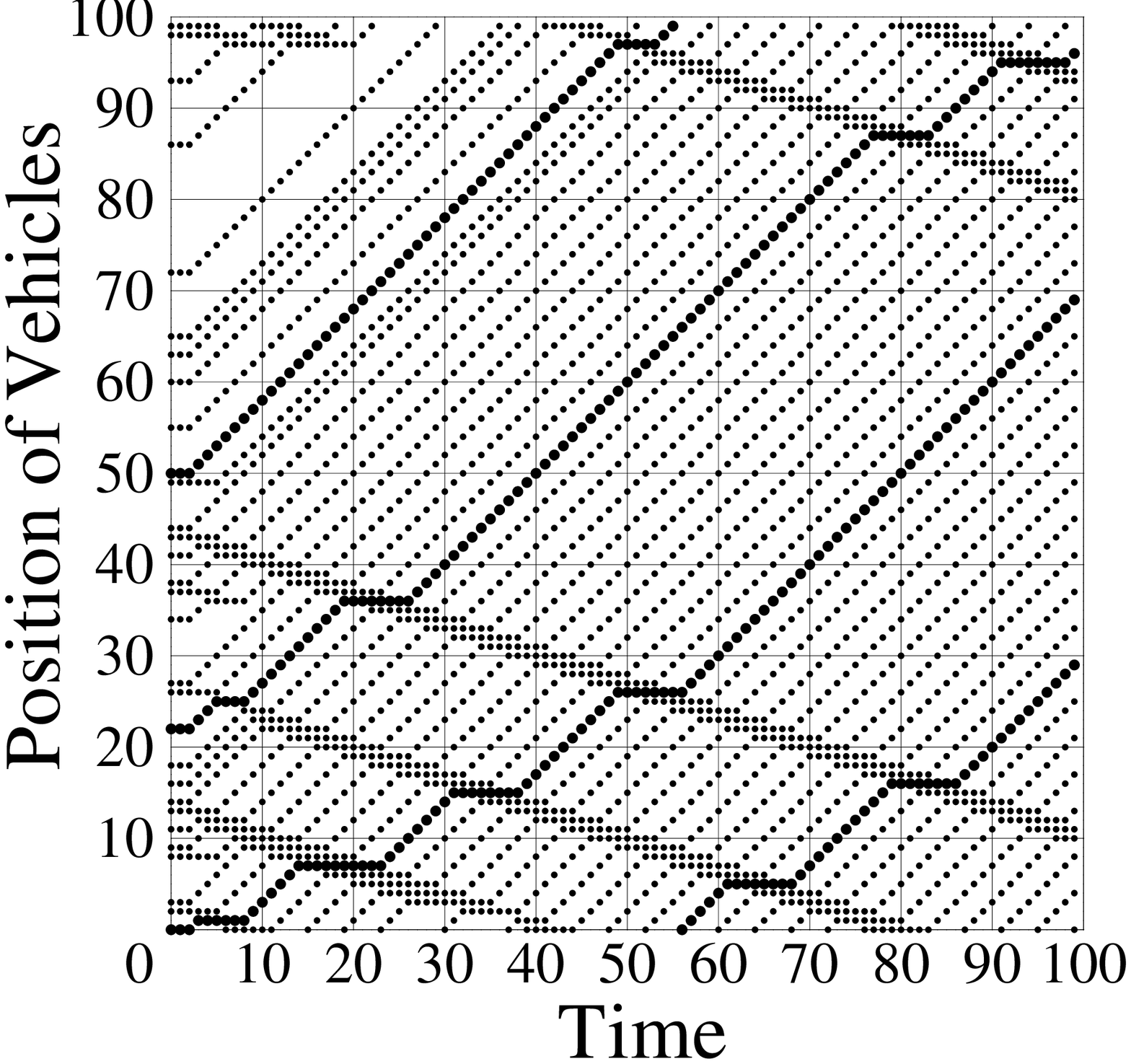}
\includegraphics[width=60mm]{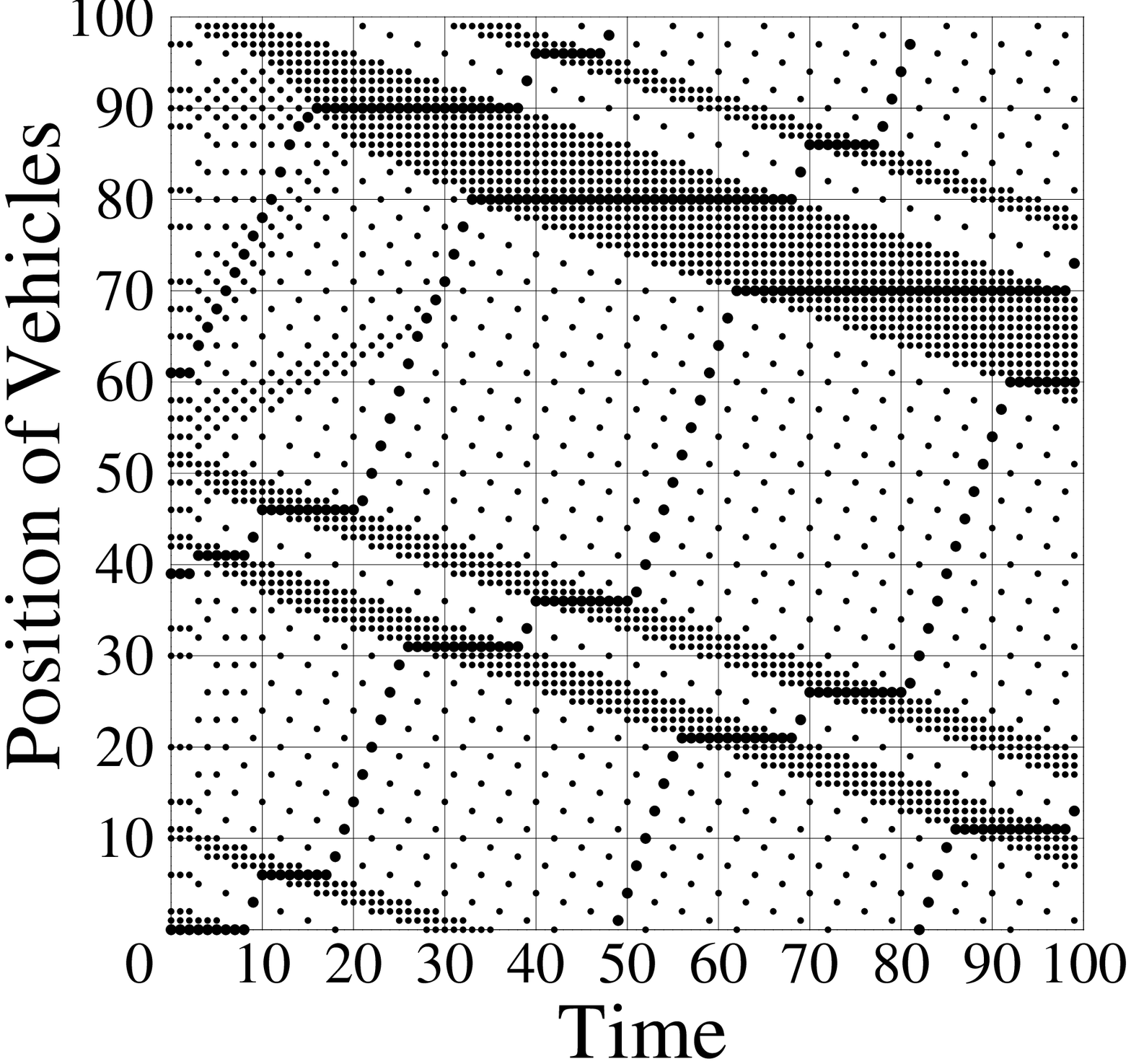}

\includegraphics[width=60mm]{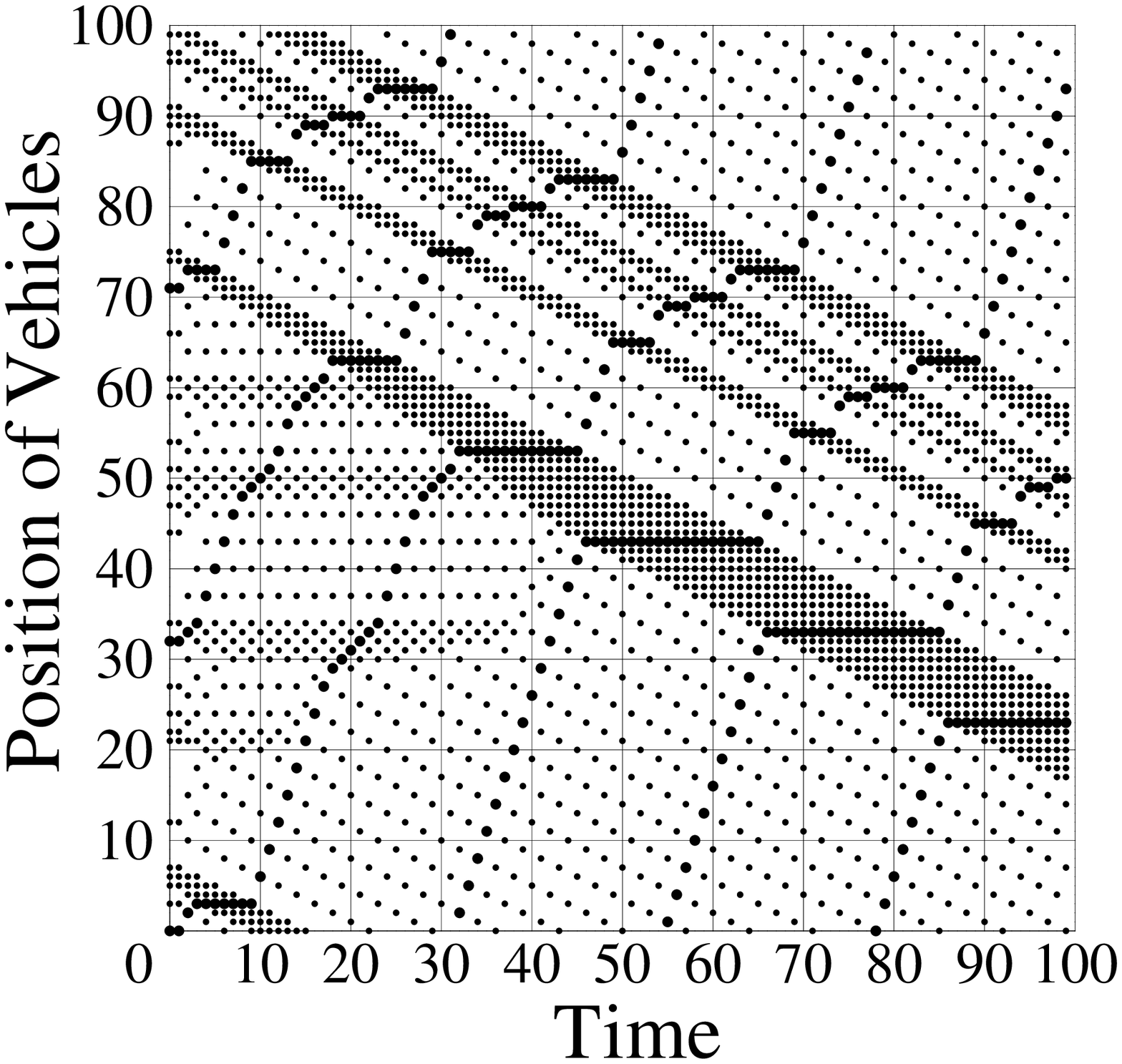}
\includegraphics[width=60mm]{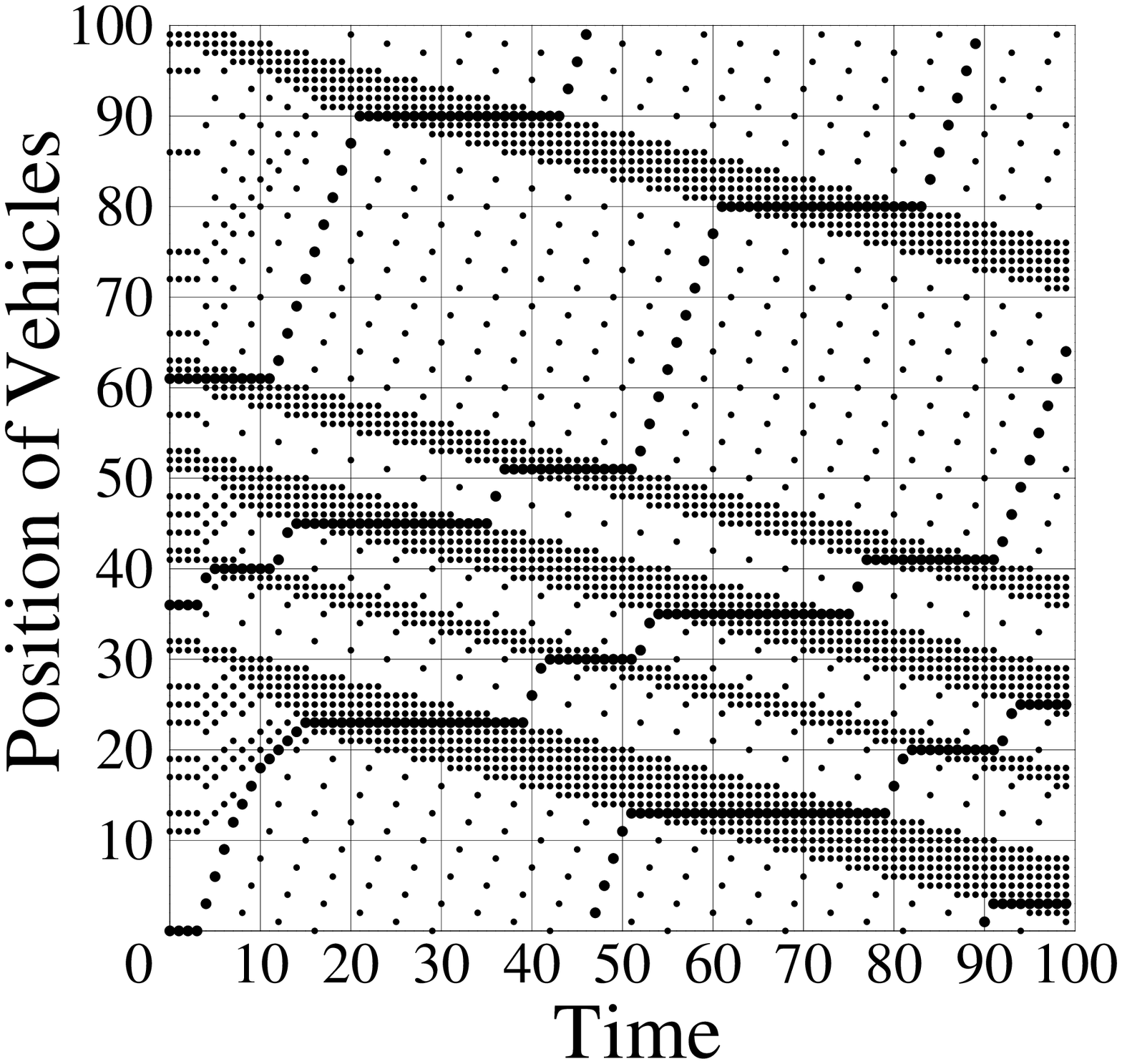}
\caption{The spatio-temporal patterns of the s2s--OV CA.
For all four patterns, the number of cars $K$ is fixed at $K=30$ 
The maximum velocities $v_0\delta t$ and the monitoring
periods $n_0$ for these patterns are
(top left) $v_0\delta t=1$, $n_0=2$,
(top right) $v_0\delta t=3$, $n_0=2$,
(bottom left) $v_0\delta t=2$, $n_0=1$ and
(bottom right) $v_0\delta t=2$, $n_0=3$, respectively.}
\label{fig:1}
\end{figure}

Spatio-temporal patterns showing 
trajectories of each vehicle are given in fig.~\ref{fig:1}. 
We choose the parameters and initial conditions so that
jams appear in the trajectories.
The two figures in the top share the same monitoring period
$n_0=2$ but their maximum velocities are different.
The top left trajectories show that the velocities of the vehicles
are zero or one, which is less than or equal to its maximum velocity
$v_0\delta t=1$. In the top right trajectories whose maximum velocity
$v_0\delta t=3$, on the other hand, 
the velocities of the vehicles read zero, one, two and three. Thus we
notice that the vehicles driven by the s2s--OV CA can run at 
any allowed integral velocity which is less than or equal to its maximum
velocity $v_0\delta t$. 

The other two figures in the bottom in fig.~\ref{fig:1}
share the same maximum velocity $v_0\delta t=2$, but their
monitoring periods are different. As is observed
in the bottom two figures, the longer the monitoring period is,
the longer it takes for the cars to get out of the traffic jam.
\begin{figure}[h]
\psfrag{stream}{direction of the stream}
\psfrag{front}{jam front}
\psfrag{t}{time}
\psfrag{x}{$x_0(=1)$}
\psfrag{n}{$\bigl(n_0(=3)+1\bigr)\delta t(=4)$}
\psfrag{0}{0}
\psfrag{1}{1}
\psfrag{2}{2}
\psfrag{3}{3}
\psfrag{4}{4}
\psfrag{5}{5}
\includegraphics[width=100mm]{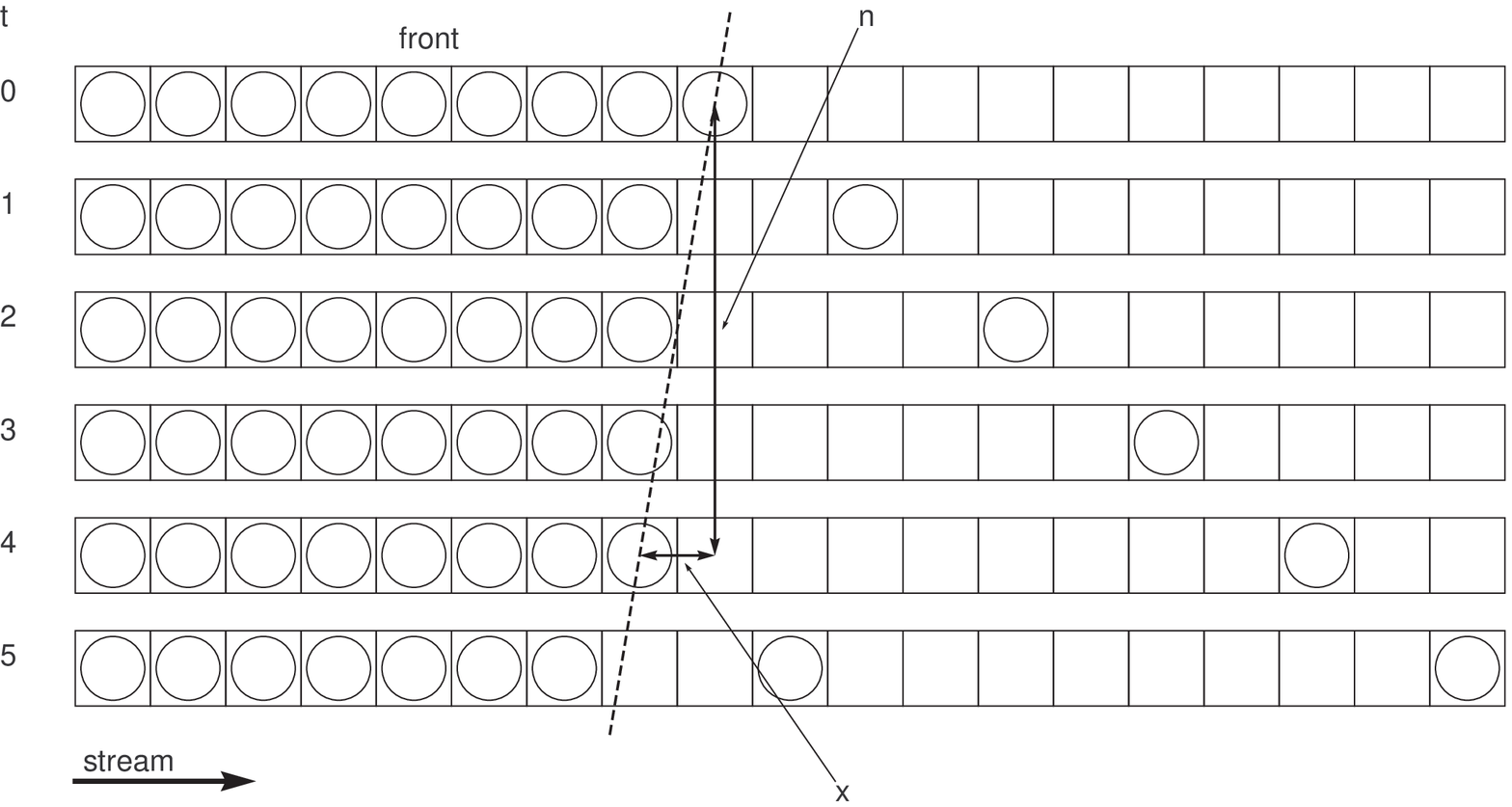}
\caption{Backward propagation of the jam front at constant
velocity $\frac{x_0}{(n_0+1)\delta t}=\frac{1}{4}$
for the case $v_0\delta t=2$, $n_0=3$ and $x_0=1$.}
\label{fig:2}
\end{figure}
The jam front is observed to propagate 
against the stream of vehicles at constant 
velocity $\frac{x_0}{(n_0+1)\delta t}$, since cars have to
wait $n_0+1$ time-steps to restart after their preceding cars restarted,
as is depicted in fig.~\ref{fig:2}.

\begin{figure}[h]
\psfrag{F}{\small Flow}
\psfrag{l}{}
\psfrag{o}{}
\psfrag{w}{}
\psfrag{D}{\small Density}
\psfrag{e}{}
\psfrag{n}{}
\psfrag{s}{}
\psfrag{i}{}
\psfrag{t}{}
\psfrag{y}{}
\psfrag{.}{\tiny .}
\psfrag{9}{\tiny 9}
\psfrag{8}{\tiny 8}
\psfrag{7}{\tiny 7}
\psfrag{6}{\tiny 6}
\psfrag{5}{\tiny 5}
\psfrag{4}{\tiny 4}
\psfrag{3}{\tiny 3}
\psfrag{2}{\tiny 2}
\psfrag{1}{\tiny 1}
\psfrag{0}{\tiny 0}
\includegraphics[width=60mm]{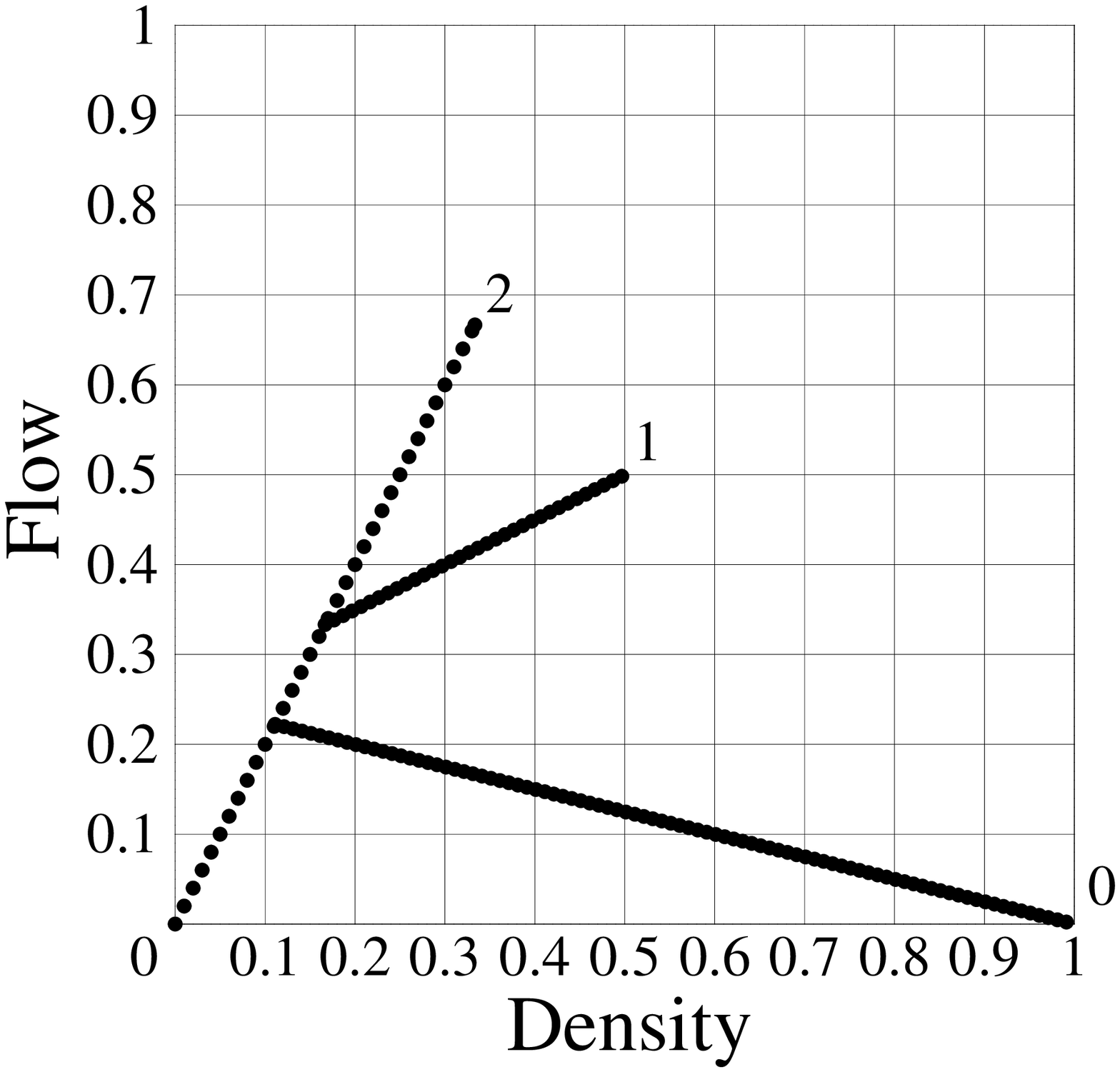}
\includegraphics[width=60mm]{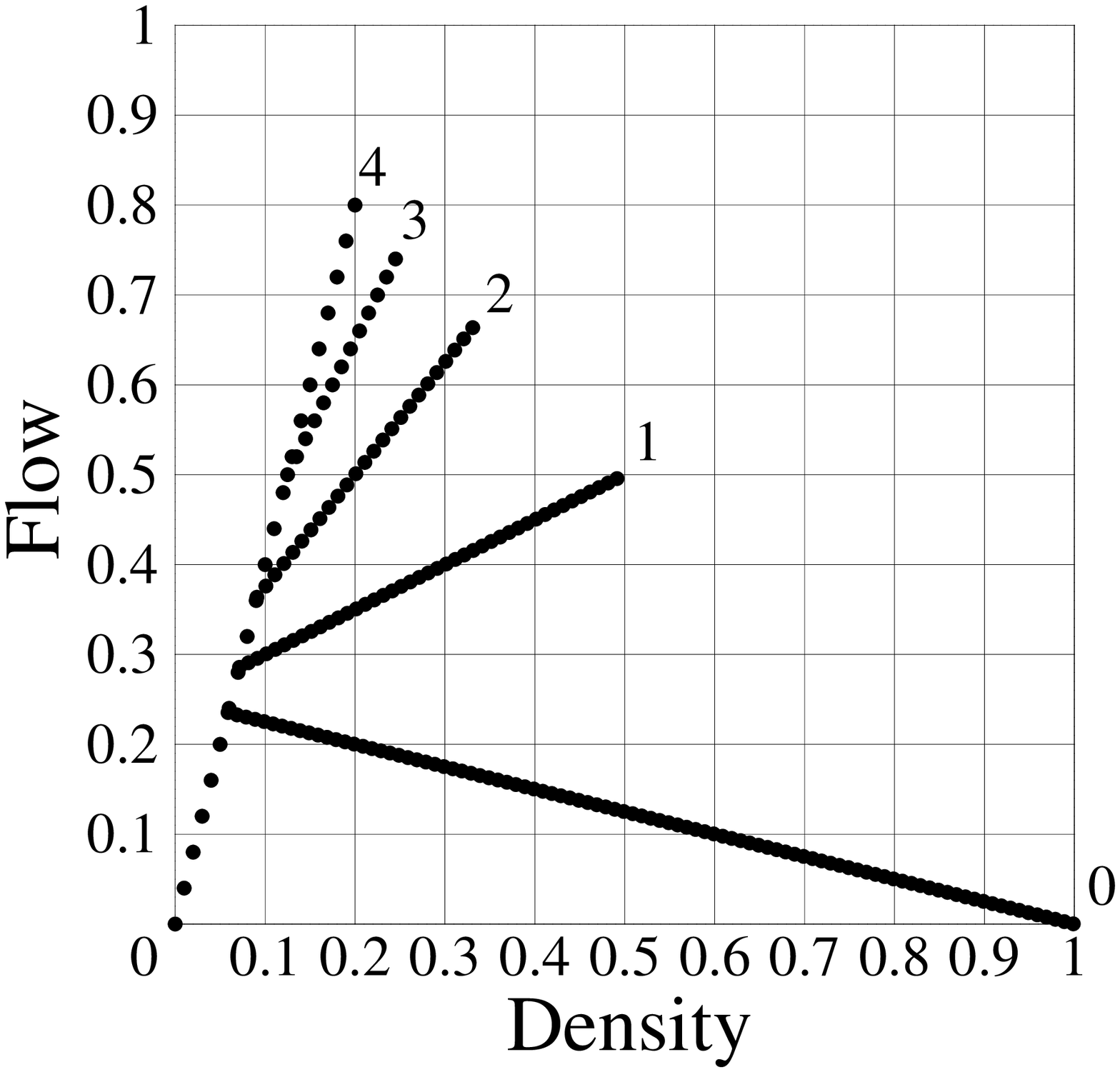}

\includegraphics[width=60mm]{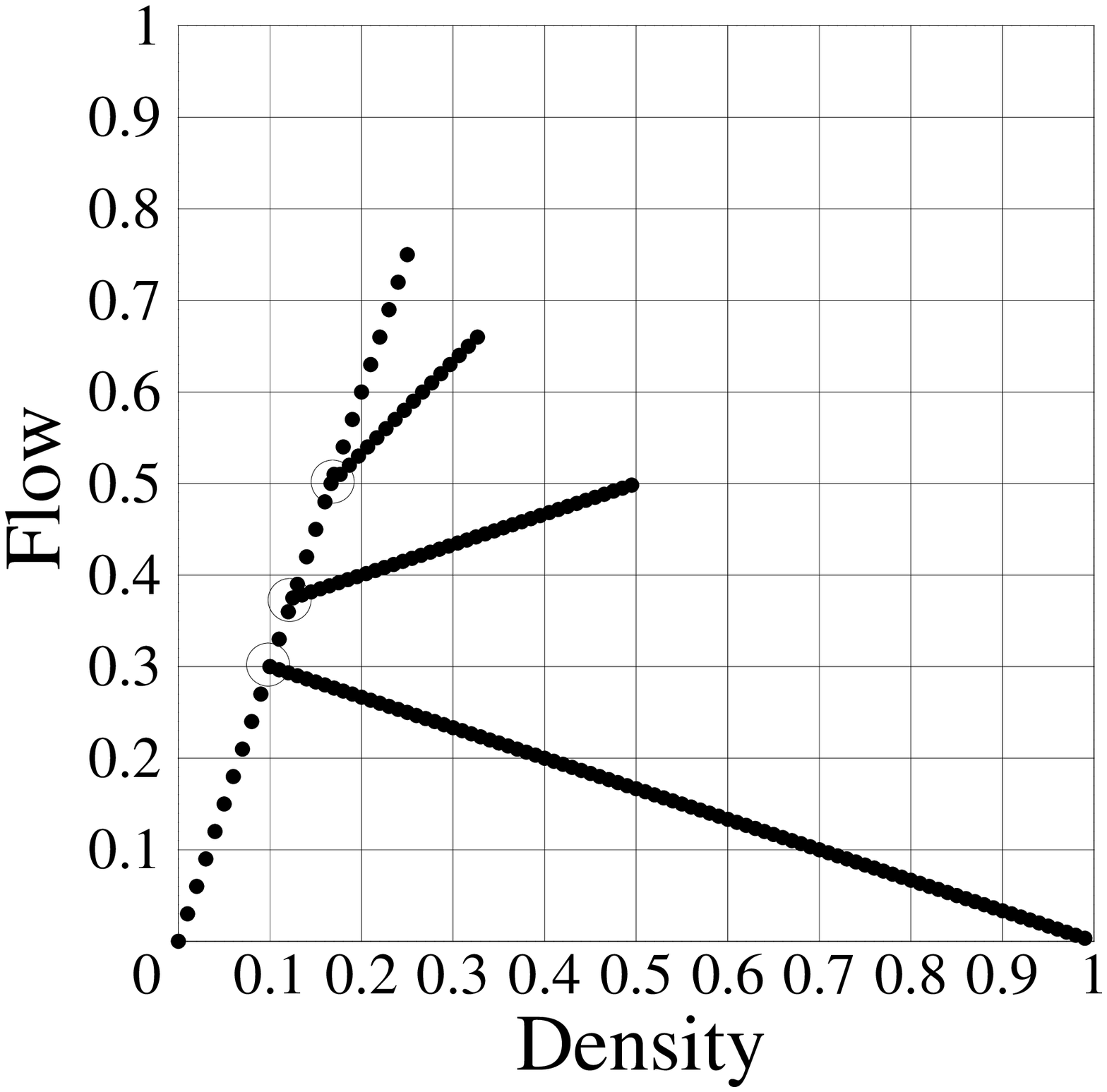}
\includegraphics[width=60mm]{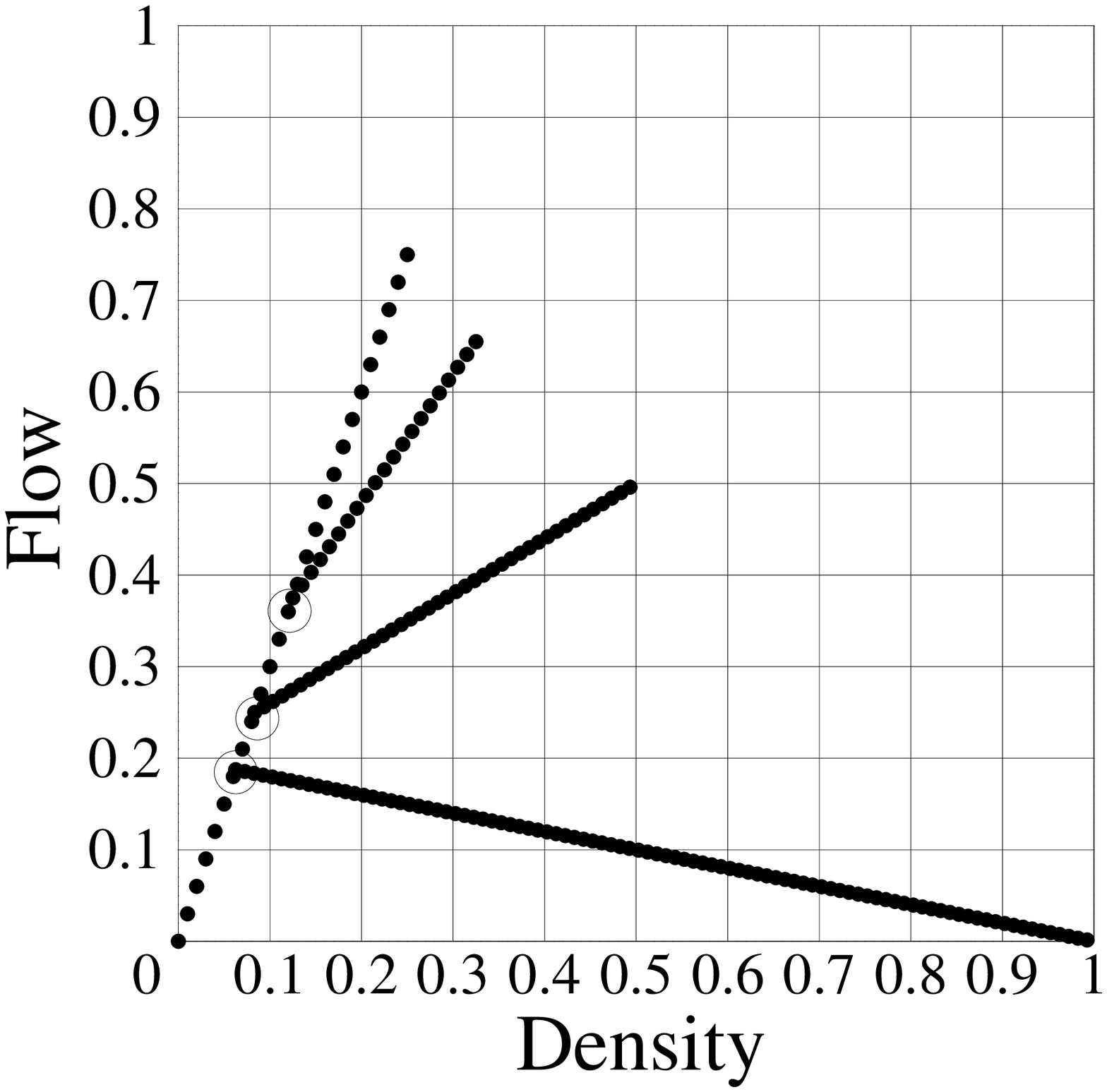}
\caption{The fundamental diagrams of the s2s--OV CA.
The flows $Q$ are computed by averaging over the time
period $800\leq n\leq 1000$. 
The maximum velocities $v_0\delta t$ and the monitoring
periods $n_0$ for these patterns are
(top left) $v_0\delta t=2$, $n_0=3$,
(top right) $v_0\delta t=4$, $n_0=3$,
(bottom left) $v_0\delta t=3$, $n_0=2$ and
(bottom right) $v_0\delta t=3$, $n_0=4$, respectively.
The inclination of the free line equals to the maximum velocity
$v_0\delta t$. The jamming line has a negative inclination.}
\label{fig:3}
\end{figure}

Fig.~\ref{fig:3} shows fundamental diagrams giving the relation
between the vehicle flow
\[
  Q:=\dfrac{1}{(n_1-n_0+1)L}\sum_{k=1}^K\sum_{n=n_0}^{n_1}
  \dfrac{x_k^{n+1}-x_k^n}{\delta t},
\]
which is equivalent to the total momentum of
vehicles per unit length,
and the vehicle density $\rho:=\frac{K}{L}$, where
$K$ is the number of vehicles. The fundamental diagrams
clearly show phase transitions from free to jam phases
as well as metastable states, which are also observed
in empirical flow-density relations~\cite{Chowdhury2000,Helbing2001}.
It is remarkable that the fundamental diagrams have
multiple metastable branches. This feature is similar to that
reported by Nishinari {\it et al.}~\cite{Nishinari2004}
By observation, we note that each fundamental diagram has
$v_0\delta t$ metastable branches and a jamming line.
The branches and the jamming line correspond to integral
velocities that are less than or equal to the maximum velocity
$v_0\delta t$. Let us confirm it with fig.~\ref{fig:3}.
The top two figures share the same monitoring
period $n_0=3$, but their maximum velocities are different.
The top left diagram corresponding to $v_0\delta t=2$ has
three branches. This number equals to that of all the
integral velocities, two, one and zero, as is depicted in the diagram.
The number of the metastable branches in the top right diagram
as well as those of the bottom two are explained in the same manner.
This observation also suggests that the monitoring period 
is irrelevant to the number of metastable branches.

All the end points of the branches 
as well as the jamming line are on the line
$\rho+Q(=\rho+Q\frac{\delta t}{x_0})=1$. This is because
the density at the end point is the maximum density $\rho_{\rm max}(v)$
that allows the velocity of the slowest 
car to be $v\delta t$. The maximum density 
$\rho_{\rm max}(v)$ is determined by
\[
  \rho_{\rm max}(v)=\dfrac{x_0}{v\delta t+x_0}.
\]
Since all the cars flow at the velocity $v\delta t$ 
when $\rho=\rho_{\rm max}(v)$, the corresponding flow is given by
$Q(\rho_{\rm max})=v\rho_{\rm max}$. Thus the relation
$\rho_{\rm max}+Q(\rho_{\rm max})\frac{\delta t}{x_0}=1$ holds.

The free line is a branch whose 
inclination equals to the maximal velocity $v_0\delta t$.
Any other metastable branch and the jamming line
branch out from the free line. By observation,
the density of the branch point of the branch corresponding
to the velocity $v\delta t$ reads
\[
  \rho_{\rm b}=\dfrac{x_0}{(v_0\delta t-v\delta t)n_0+v_0\delta t +x_0}.
\]
This observation is explained as follows. Suppose 
one car, say the car $k$, 
runs at the velocity $v$ and the other $K-1$ cars
run at the maximum velocity $v_0$.
At the moment the $k$-th car
slows down to $v$, the headway between
the cars $k$ and $k+1$ is $v\delta t +x_0$. Since
it takes at least $n_0+1$ time-steps for the car $k$ to
speed up to $v_0$, the headway between the cars $k$ and $k+1$
expands up to $H=(v_0\delta t-v\delta t)(n_0+1)+v\delta t+x_0
=x_0/\rho_{\rm b}\geq v_0\delta t$ by the time the $k$-th
car speeds up to $v_0$. If all the cars can obtain the
headway $H$, slow cars running at the velocity $v$ disappear
in the end. Thus the density at 
the branch point of the branch corresponding
to the velocity $v\delta t$ is given by $\rho_{\rm b}=x_0/H$.
Note that the density at the branch point becomes smaller 
as the monitoring period becomes larger.

\section{Concluding remarks}
Through an extension of the ultradiscretization for 
the OV model~\cite{Takahashi2009},
we introduced the ds2s--OV~\eref{eq:ds2s--OV} and s2s--OV~\eref{eq:s2s--OV}
models as ultradiscretizable traffic flow models.
The model is a hybrid of the OV~\cite{Bando1995}
and the s2s~\cite{Takayasu1993} models
whose ultradiscrete limit gives a generalization of a special case
of the uOV model by Takahashi and Matsukidaira~\cite{Takahashi2009}.
The phase transition from free to
jam phases as well as the existence of multiple metastable states
were observed in numerically obtained fundamental diagrams 
for the s2s--OV CA~\eref{eq:s2s--OVCA}, 
which are special cases of the us2s--OV model~\eref{eq:us2s--OV}.

Detailed studies on the properties of the hybrid models~\eref{eq:ds2s--OV},
\eref{eq:s2s--OV}, \eref{eq:us2s--OV} and \eref{eq:s2s--OVCA} such as
exact solutions, comparison with other traffic flow models
as well as empirical data remain to be investigated.

\begin{acknowledgements}
The authors are grateful to D.~Takahashi, J.~Matsukidaira,
A.~Tomoeda, D.~Yanagisawa and R.~Nishi for their valuable comments
at the spring meeting of JSIAM in March, 2009.
\end{acknowledgements}

\end{document}